**Nanoscale structure and mechanism for enhanced electromechanical response of highly-strained BiFeO$_3$ thin films**

By *A. R. Damodaran*, *C.-W. Liang*, *Q. He*, *C.-Y. Peng*, *L. Chang*, *Y.-H. Chu*, and *L. W. Martin**

[*]     Prof. L. W. Martin, Dr. C.-W. Liang, A. Damodaran,
Department of Materials Science and Engineering and F. Seitz Materials Research Laboratory
University of Illinois, Urbana-Champaign
Urbana, IL 61801
E-mail: lwmartin@illinois.edu

        Dr. Y.-H. Chu, Dr. C.-W. Liang, C.-Y. Peng, L. Chang
Department of Materials Science and Engineering
National Chiao Tung University
Hsinchu, 30010 Taiwan

        Dr. Q. He
Advanced Light Source
Lawrence Berkeley National Laboratory
Berkeley, CA 94720

**The nanostructural evolution of the strain-induced structural phase transition in BiFeO$_3$ is examined. Using high-resolution X-ray diffraction and scanning-probe microscopy-based studies we have uniquely identified and examined the numerous phases present at these phase boundaries and have discovered an intermediate monoclinic phase in addition to the previously observed rhombohedral- and tetragonal-like phases. Further analysis has determined that the so-called mixed-phase regions of these films are not mixtures of rhombohedral- and tetragonal-like phases, but intimate mixtures of highly-distorted monoclinic phases with no evidence for the presence of the rhombohedral-like parent phase. Finally, we propose a mechanism for the enhanced electromechanical response in these films including how these phases interact at the nanoscale to produce large surface strains.**

Piezoelectric materials have been used in a wide array of technologies[1] and are potential candidates for energy harvesting and memory applications.[2,3] State-of-the-art piezoelectrics are based on perovskite materials with complex chemistries including $PbZr_xTi_{1-x}O_3$ (PZT) at x = 0.52 and $(1-x)Pb(Mg_{1/3}Nb_{2/3})O_3$-$xPbTiO_3$ (PMN-PT) at x = 0.33. Special attention has been given to these materials due to the presence of morphotropic phase boundaries (MPB) – or a nearly temperature insensitive, compositionally driven boundary between tetragonal- and rhombohedral-polymorphs which gives rise to enhanced piezoelectricity.[4] Recently, it has been observed that MPB-like features can also exist in appropriately strained, chemically simple materials.[5,6,7] For instance, it has been shown that $BiFeO_3$, which possesses a strong electrical polarization (~90-100 $\mu C/cm^2$) and is a candidate for the replacement of lead-based ferroelectrics,[8,9] can exhibit a strain-induced structural phase transition which gives rise to enhanced electromechanical response.[7] Using epitaxial thin film growth, X-ray reciprocal space mapping (RSM), atomic (AFM) and piezoresponse (PFM) force microscopy we report the nanoscale structural evolution of such strain-induced phase boundaries in $BiFeO_3$. We report on the true structure of these phase boundaries – including the observation of never before reported phases, the nanoscale spatial location of these phases, as well as the interplay between these phases and how this results in strong electromechanical actuation in these materials. These concepts represent unique advanced in our understanding of this technologically relevant material and are extremely timely.

$BiFeO_3$ is a rhombohedrally-distorted (*R3c*) multiferroic perovskite exhibiting antiferromagnetism that is coupled with ferroelectric order.[10,11] Although the structure of $BiFeO_3$ had been studied for many years,[12,13,14] in 2005 the structural stability of the parent phase had come into question.[15,16] An exhaustive high-temperature neutron-diffraction study of bulk $BiFeO_3$ powders examined the complex structure and phases in this material and reported that both the β- and γ-phases are orthorhombic and that there is no evidence for a monoclinic phase in unstrained $BiFeO_3$.[17] Other thin film studies, however, have shown that a tetragonally-distorted phase (derived from a structure with *P4mm* symmetry, *a* ~ 3.665 Å, and *c* ~ 4.655 Å) with a large spontaneous polarization may be possible.[15,18,19] Furthermore, so-called mixed-phase thin-films possessing tetragonal- and rhombohedral-like phases in complex stripe-like structures that give rise to enhanced electromechanical responses were also reported.[7] Since this report, additional information has come forth about these materials including the fact that the tetragonal-like phase is actually monoclinically distorted (possessing *Cc, Cm, Pm, or Pc* symmetry).[20,21, 22,23] Nonetheless, a thorough understanding of the complex structure of these phase boundaries in $BiFeO_3$ remains incomplete. Building off of our understanding of structure and response in chemically driven MPB systems,[24] the current work uniquely identifies the boundary nanostructure and the pathway for enhanced electromechanical response in $BiFeO_3$ films.

Epitaxial $BiFeO_3$ thin films of thickness 20-200 nm were deposited by pulsed laser deposition from a $Bi_{1.1}FeO_x$ target at 700°C in oxygen pressures of 100 mTorr on single-crystal $LaAlO_3$ (001) substrates and were cooled at 760 Torr. Detailed structural information was obtained using high-resolution RSM (X'Pert MRD Pro, PIXcel detector, Panalytical). The film topography was studied with AFM and piezoelectric, ferroelectric, and switching studies were completed using PFM (Cypher, Asylum Research).

Consistent with prior reports,[7] films less than ~40 nm thick exhibited only the distorted tetragonal-like phase. The complexity of this system requires a rigorous definition of the phases present, thus throughout the remainder of this manuscript, we will refer to this monoclinically distorted tetragonal-like phase as the $M_{II}$-phase. AFM of these samples reveals atomic-level terraces with step-heights corresponding to single unit cells (~4.65Å) (**Fig. 1A**). Detailed RSM of the 001-diffraction condition (**Fig. 1B**) reveals the presence of only two peaks – the 001-diffraction peak for the $LaAlO_3$ substrate and that for the $M_{II}$-phase of $BiFeO_3$. As film thickness is increased, shown here is data for a 130 nm thick film, regions of pure $M_{II}$-phase (bright areas, **Fig. 1C**) and mixed-phase regions (striped areas, Fig. 1C) are observed. Early reports suggested that these mixed-phase regions were an intimate mixture of the tetragonal- and rhombohedral-like phases. In turn, we will provide evidence of a more complex nanostructural arrangement. We will henceforth refer to the monoclinically distorted rhombohedral bulk-like phase as the R-phase. Both our results and the results of previous works focusing on this R-phase (Ref. [25, 26, 27]) reveal similar lattice parameters and monoclinic distortions. RSM of the 001-diffraction condition of this film (**Fig. 1D**) reveals significant complexity. Peaks corresponding to the 001-diffraction peak of the substrate, the R-phase ($c$ = 3.979 +/- 0.003 Å), and the $M_{II}$-phase ($c$ = 4.667 +/- 0.005 Å) are labeled accordingly. The peak corresponding to the R-phase has very low intensity and this peak has been found to be absent in other films. Additionally, there are two new sets of diffraction peaks. The first set is located on either side of the $M_{II}$-phase peak in $Q_x$-space at a $Q_y$-value of ~1640 (labeled as $M_{II,tilt}$, Fig. 1D). These peaks correspond to a phase that possesses the same $c$ (4.667Å) as the $M_{II}$-phase, but is tilted by ~1.53° along the [100] relative to the [001] sample normal. The second set comprises four diffraction peaks at a $Q_y$-value of ~1850 which represent a previously unobserved phase, henceforth referred to as an intermediate monoclinic phase ($M_I$-phase). The $M_I$-phase has $c$ = 4.168Å – intermediate between the R- and $M_{II}$-phases – and exhibits two pairs of peaks – the outer pair suggests this phase is tilted ~2.88° along the [100] relative to the [001] sample normal and the inner pair suggests a rotation of ~0.4° about the sample normal. The outer- and inner-peak pairs are thought to correspond to diffraction from the same phases in the two stripe-sets (rotated by 90° in the plane of the film).

In order to better understand this complex structure, we have performed high-resolution RSM studies of a number of off-axis diffraction peaks (i.e., 103, 013, 113, 1-13, 203, and 023). RSMs of the 103- (**Fig. 2A**) and the 113- (**Fig. 2B**) diffraction peaks of a 28 nm thick film possessing only the $M_{II}$-phase reveal peak splitting into three and two distinct peaks, respectively. This splitting identifies that the $M_{II}$-phase is monoclinically distorted with a small tilt along the in-plane <100>. The RSM studies reveal that the $M_{II}$-phase exhibits anisotropic in-plane lattice parameters ($a \sim 3.74$ Å, $b \sim 3.82$ Å), $c/a \sim 1.23$, a monoclinic angle $\beta = 88.1°$, and suggest that the distorted tetragonal-like phase is indeed monoclinic with either *Pm* or *Pc* symmetry.[22] Thicker films (i.e., 130 nm) exhibiting complex mixed-phase topography provide further information. The peaks corresponding to the $M_{II}$-phase are again observed to split into three and two diffraction peaks in the 103- (**Fig. 2C**) and 113- (**Fig. 2D**) RSMs, respectively. Furthermore, we observe additional diffraction peaks corresponding to the $M_{II,tilt}$-phase which suggest that *a* and *b* closely match those of the $M_{II}$-phase and confirm that this phase is tilted from the substrate normal by ~1.53° along the <100>. Finally, analysis of the 001-, 103- (Fig. 1D, 2C), and other peaks for the $M_I$-phase has enabled us to estimate the lattice parameters for this intermediate phase to be *a* and *b* ~3.82Å and *c* ~4.168Å and to confirm that this phase is tilted from the substrate normal by ~2.88°.

In the remainder of this work we investigate the nanoscale spatial distribution of these various phases and propose a pathway for the enhanced electromechanical response. AFM and PFM studies of mixed-phase samples reveal the presence of stripe-like contrast within $M_{II}$-phase regions in the lateral (in-plane) response (**Supp**. **Fig. 1A** and **B**). This can be attributed to the monoclinic distortion reported above and is consistent with prior observations.[21,28] High-resolution AFM also provides direct images of the structural features observed in the X-ray diffraction. Evidence for regions of pure $M_{II}$-phase and mixed-phase regions (saw-tooth-like features on the surface) are observed in 140 nm thick films (**Fig. 4A**). The mixed phase regions *do not* possess a flat-bottom or well-like structure, but are saw-tooth in nature as demonstrated by the indicative line-trace (blue-line in Fig. 4A, data **Fig. 4B**). We have found that these features are asymmetric and, in this example, the left side of the saw-tooth is inclined to the substrate surface with an angle of ~1.8°, while the right side of the saw-tooth is inclined to the surface at an angle of ~2.8° This suggests that the phases observed in the X-ray diffraction studies to be tilted by ~ 1.6° ($M_{II,tilt}$-phase) and ~2.8° ($M_I$-phase) are located in intimate proximity to one another in the mixed-phase, stripe-regions (additional data provided in **Supp. Fig. 2**). Prior work suggested an intimate mixture of the R- and $M_{II}$-phases in these regions, but our data indicates otherwise. Further evidence against this intimate mixture of the R- and $M_{II}$-phases is provided in the line-trace parallel to the stripe-like features (red line Fig 4A, data **Fig. 4C**) that shows that there is an overall depression of the surface height of ~5 nm in the mixed-phase regions. Such surface depressions are consistent with the presence of the $M_I$- and $M_{II,tilt}$-phases only (see Supporting Information for further details). Studies of films from 20-200 nm reveal no appreciable change in the X-ray peak intensity of the R-phase throughout this thickness range, suggesting that very small fractions of the film are made up of the R-phase. A similar lack of the R-phase diffraction peak was observed previously (see Fig. 1 from Ref. 7). These results thus lead us to conclude that the mixed-phase regions are made up not of an intimate mixture of R- and $M_{II}$-phases, but an intimate mixture of highly distorted (or tilted), monoclinic phases – in particular the $M_I$- and $M_{II,tilt}$-phases.

Application of electric fields to these mixed-phase samples results in strong electromechanical responses. Local switching studies using PFM reveal surface electromechanical strains as large as 4-5% (**Fig. 4**). In order to better understand the mechanism for such large electromechanical responses, we studied the evolution of the surface structure and polarization of a 1 x 1 μm region of a mixed-phase 100 nm $BiFeO_3$/5 nm $La_{0.5}Sr_{0.5}CoO_3$/$LaAlO_3$ (001) thin film by applying different voltages to the region within the red box in Fig. 5A using the PFM. After each applied voltage, we captured both the topography (left images **Fig. 5**) and the out-of-plane orientation of the polarization (right images Fig. 5). In the as-grown configuration, the film possesses a mixed-phase structure and is entirely down polarized (Fig. 5A). Upon applying a 5.25V potential, we have observed no change in the surface structure or out-of-plane polarization (**Fig. 5B**). Upon further increasing the applied potential to 10.25V, however, we observe the ability to electrically drive the mixed-phase region to be entirely made-up of the $M_{II}$-phase (**Fig. 5C**). If we then begin to drive the same region with small negative potentials (e.g., -3V) we observe the ability to completely and reversibly switch the material back into a mixed-phase structure similar to the as-grown configuration (**Fig. 5D**). Further increasing the negative potential to -5.25 V results in ferroelectrically switching the material (**Fig. 5E**), but maintains the mixed-phase structure. Application of large negative biases (-9V) results in the eventual transformation of the mixed-phase structure into an upward-poled version of the $M_{II}$-phase (**Fig. 5F**). The polarization of this $M_{II}$-phase is pointing opposite to that in the structure shown in Fig. 5C. If we now apply positive potentials to this region we can return to the mixed-phase structure similar to the as-grown state, but with a reversed out-of-plane polarization (**Fig. 5G**). Finally, by further increasing to a positive potential of 5.25V we return the structure and polarization to the as-grown configuration (**Fig. 5H**).

By navigating the hysteretic nature of electric field response in this material, we have learned a number of important features of this system. First, large surface strains of generally 4-5% occur any time the material transforms form a mixed-phase structure to the $M_{II}$-phase. Second, these transformations between these two states are entirely reversible. Third, there are numerous pathways to achieve large electromechanical responses in these materials. The first, as described in Figs. 5A-H and in the corresponding schematic hysteresis loop in **Fig. 5I** occurs upon traversing the entire hysteresis loop. Following the path from A) to H) in Fig. 5I, we undergo four transformations that give rise to large surface strains. Traversing this major hysteresis loop requires the input of considerable energy and two ferroelectric polarization switching events. On the other hand, we can achieve the same large electromechanical responses *without the need to ferroelectrically switch the sample*. By following the minor hysteresis loops B-C-D-B or E-F-G-E (one is highlighted in red in Fig. 5I) we can repeatedly transform between the mixed-phase structure and the $M_{II}$-phase, thereby obtaining large electromechanical responses without switching the polarization direction. This is an exciting discovery and could have important implications for utilization of these materials in devices – including possible low-power applications.

Finally, we propose a mechanism for this enhanced electromechanical response. The key appears to be the ability to transform from the $M_I$-phase to the $M_{II,tilt}$-phase to the $M_{II}$-phase through a diffusion-less phase transition. Focusing on the minor hysteresis loop B-C-D-B and our step-by-step probing of the material, it appears that the application of a positive potential drives the preferential growth of the $M_{II,tilt}$-phase at the expense of the $M_I$-phase in the mixed-phase regions (**Fig. 5J**) until a purely $M_{II,tilt}$ region is obtained. At this point the material undergoes a final rotation of the polarization towards the surface normal and the development of the $M_{II}$-phase which has preferential alignment of polarization relative to the applied field direction. We note that we cannot uniquely identify if this polarization rotation occurs only after the material has transformed to the $M_{II}$-phase or if it gradually happens during the transformation. From a geometrical argument, however, one might expect gradual rotation of the polarization during the growth of the $M_{II,tilt}$-phase as the tilt is no longer needed to accommodate the $M_I$-phase and the elastic strain. Regardless, at this point the $M_{II}$-phase is stable and only upon application of negative potentials does the sample undergo a reversal of the transformation and return to the as-grown structure. Such a hysteresis loop provides the actuation and deactuation of the large surface displacement required for applications.

In conclusion, these results have shown that the added complexity associated with the strain-induced structural phase boundaries in $BiFeO_3$ films are key in determining the response of these materials to applied fields. We have observed that the films possess a number of phases including the $M_I$-, $M_{II}$-, and $M_{II,tilt}$-phases, that the mixed-phase regions consist of an intimate mixture of two highly-tilted, monoclinic phases ($M_I$- and $M_{II,tilt}$-phases), and have provided proposed mechanism to describe how the material can easily transform between structural polymorphs under applied electric-fields and produce strong electromechanical responses. Such unique behavior has significant implications for numerous applications and could help shed light on similar boundaries in other high-performance piezoelectrics.

*Acknowledgements*

The authors would like to acknowledge the help and scientific insights of Dr. S. MacLaren and Dr. M. Sardela at the Center for Microanalysis of Materials at UIUC. The work at UIUC was supported by the Army Research Office under grant W911NF-10-1-0482 and by Samsung Electronics Co., Ltd. under grant 919 Samsung 2010-06795. Experiments at UIUC were carried out in part in the Frederick Seitz Materials Research Laboratory Central Facilities, which are partially supported by the U.S. Department of Energy under grants DE-FG02-07ER46453 and DE-FG02-07ER46471. The work at National Chiao Tung University was supported by the National Science Council under contract No. NSC-099-2811-M-009-003.

Received: ((will be filled in by the editorial staff))
Revised: ((will be filled in by the editorial staff))
Published online: ((will be filled in by the editorial staff))

**Figure Captions**

**Figure 1.** AFM image and RSM of the 001-diffraction peak for 28 nm (A and B, respectively) and a 130 nm thick (C and D, respectively) BiFeO$_3$/LaAlO$_3$ (001) thin films. Note the presence of Kiessig fringes in part B.

**Figure 2.** RSMs of the 103- and 113-diffraction conditions of a 28 nm (A and B, respectively) and 130 nm (C and D, respectively) thick BiFeO$_3$/LaAlO$_3$ (001) film.

**Figure 3.** A) AFM topography, B) line-trace along the blue line in part A), and C) line-trace along red line in part A) of a 140 nm thick BiFeO$_3$/LaAlO$_3$ (001) thin film.

**Figure 4** AFM images of 100 nm and 50 nm BiFeO$_3$/La$_{0.5}$Sr$_{0.5}$CoO$_3$/LaAlO$_3$ (001) heterostructures in the A) and D) as-grown state, B) and E) following electrical poling, and C) and D) corresponding line traces at the red-lines.

**Figure 5.** AFM image (left) and vertical PFM image (right) of 100 nm BiFeO$_3$/La$_{0.5}$Sr$_{0.5}$CoO$_3$/LaAlO$_3$ (001) in the A) as-grown state and after being poled in the red box at B) 5.25 V, C) 10.25 V, D) -3 V, E) -5.25 V, F) -9 V, G) 4.5 V, and H) 5.25 V. (All images are 1 x 1 μm). I) A schematic hysteresis loop with letters corresponding to the images in A)-H) shows the multiple pathways to enhanced electromechanical response. J) Illustration of the proposed mechanism for the large electromechanical response without the need for ferroelectric switching.

**Figure 1**

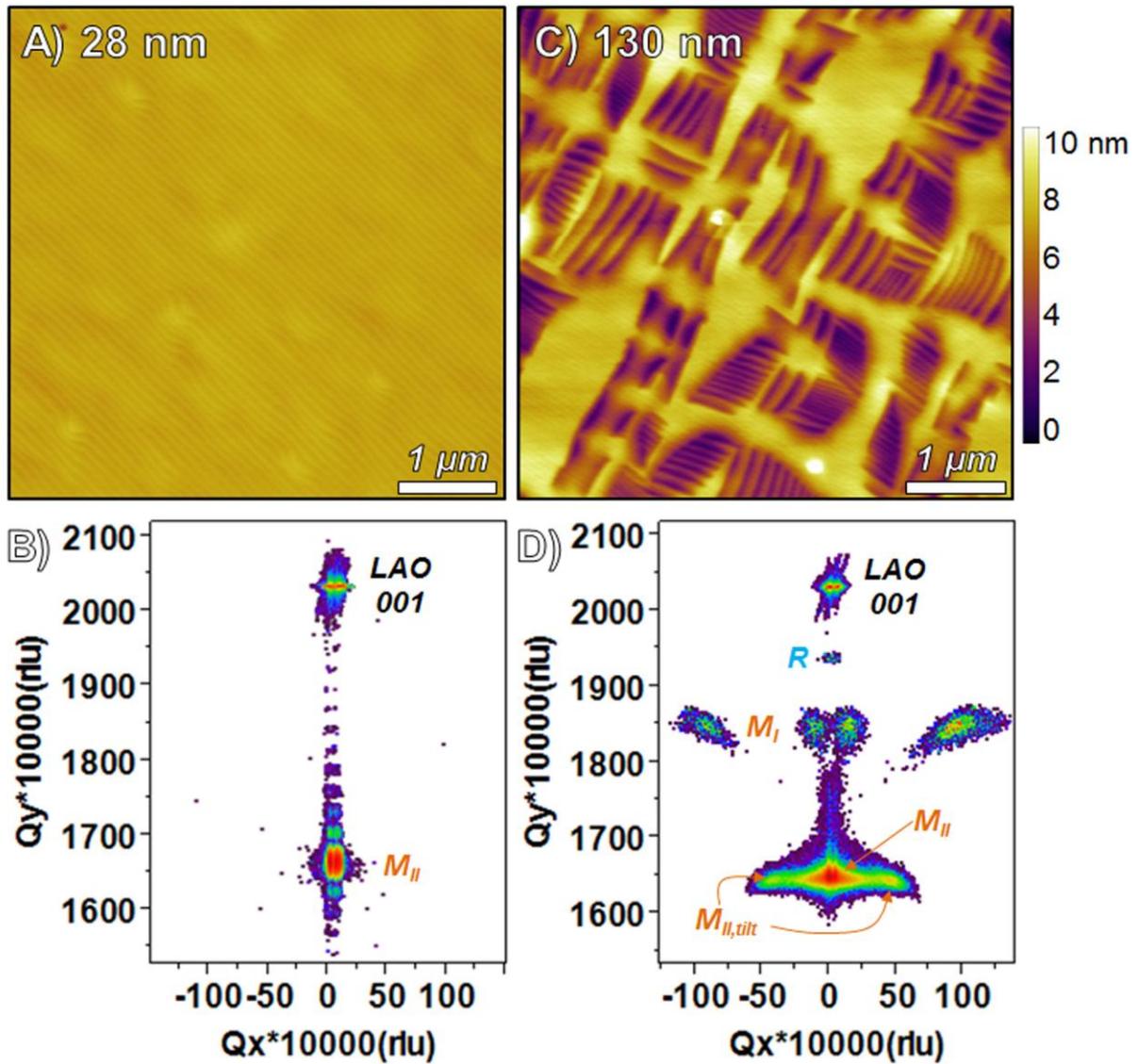

**Figure 1.** AFM image and RSM of the 001-diffraction peak for 28 nm (A and B, respectively) and a 130 nm thick (C and D, respectively) $BiFeO_3/LaAlO_3$ (001) thin films. Note the presence of Kiessig fringes in part B.

**Figure 2**

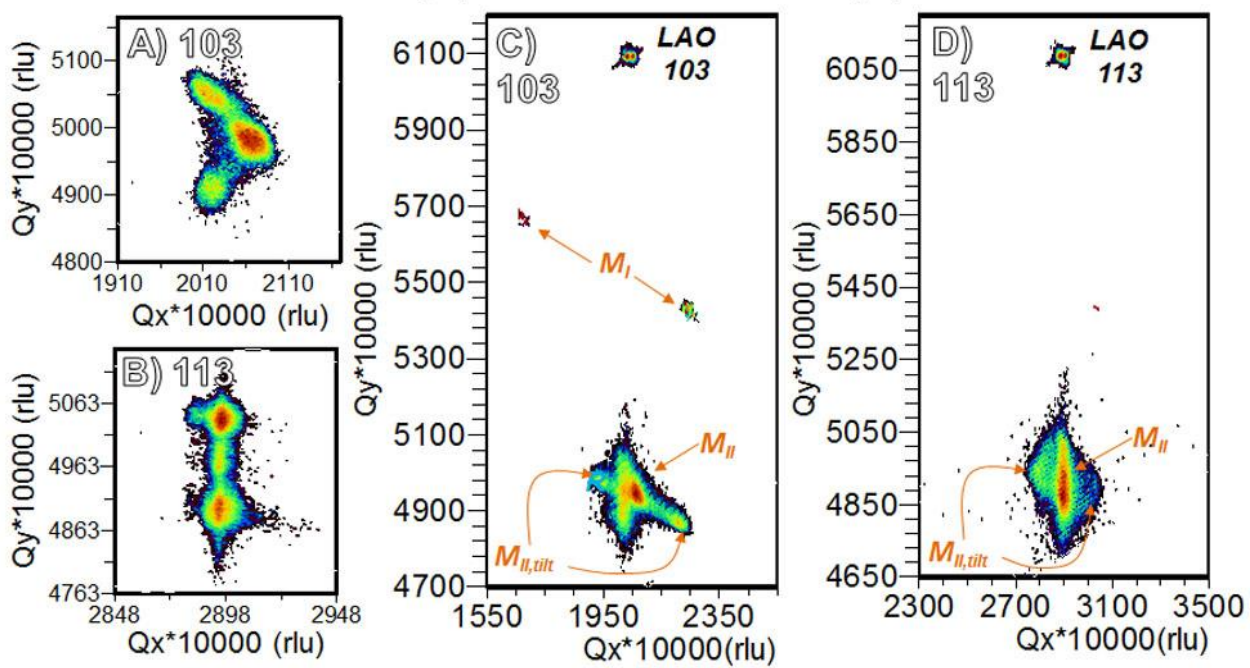

**Figure 2.** RSMs of the 103- and 113-diffraction conditions of a 28 nm (A and B, respectively) and 130 nm (C and D, respectively) thick $BiFeO_3/LaAlO_3$ (001) film.

**Figure 3**

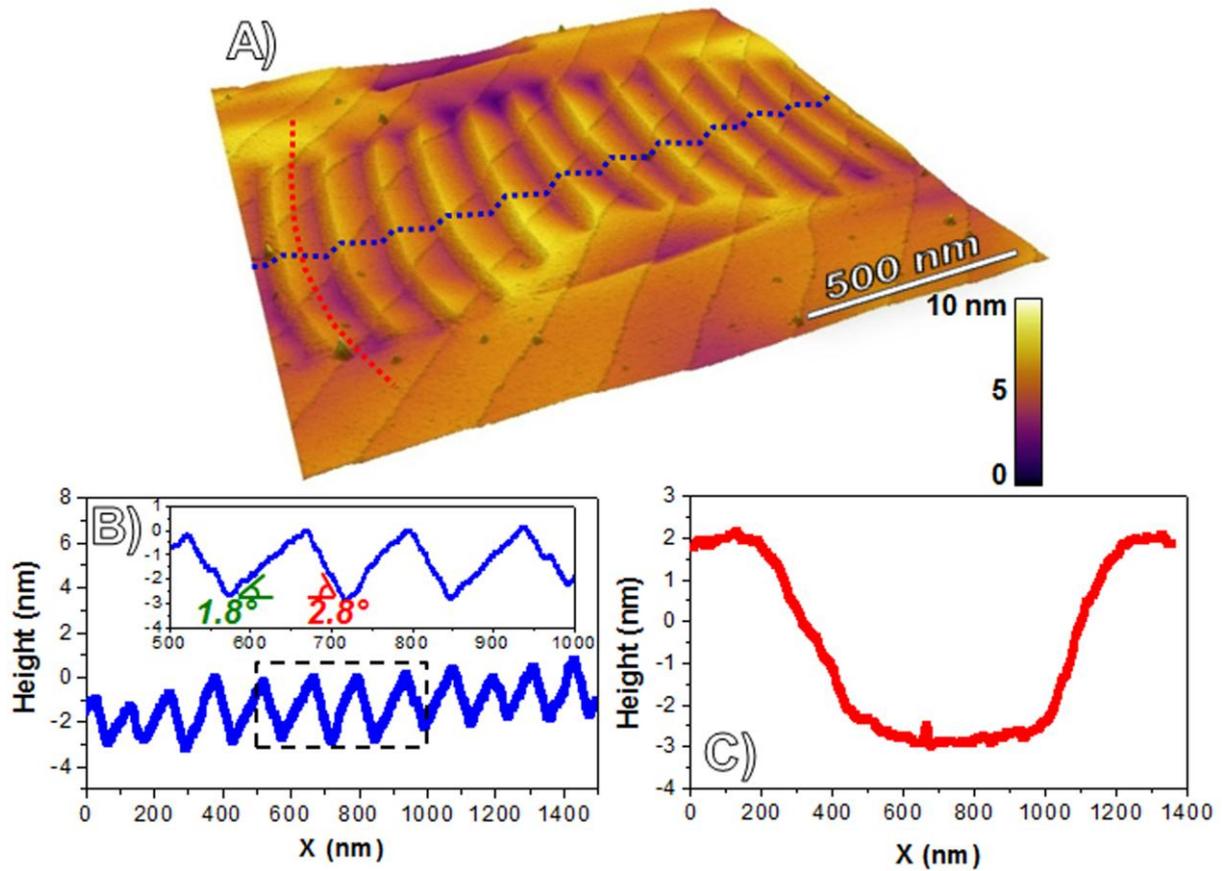

**Figure 3.** A) AFM topography, B) line-trace along the blue line in part A), and C) line-trace along red line in part A) of a 140 nm thick $BiFeO_3/LaAlO_3$ (001) thin film.

**Figure 4**

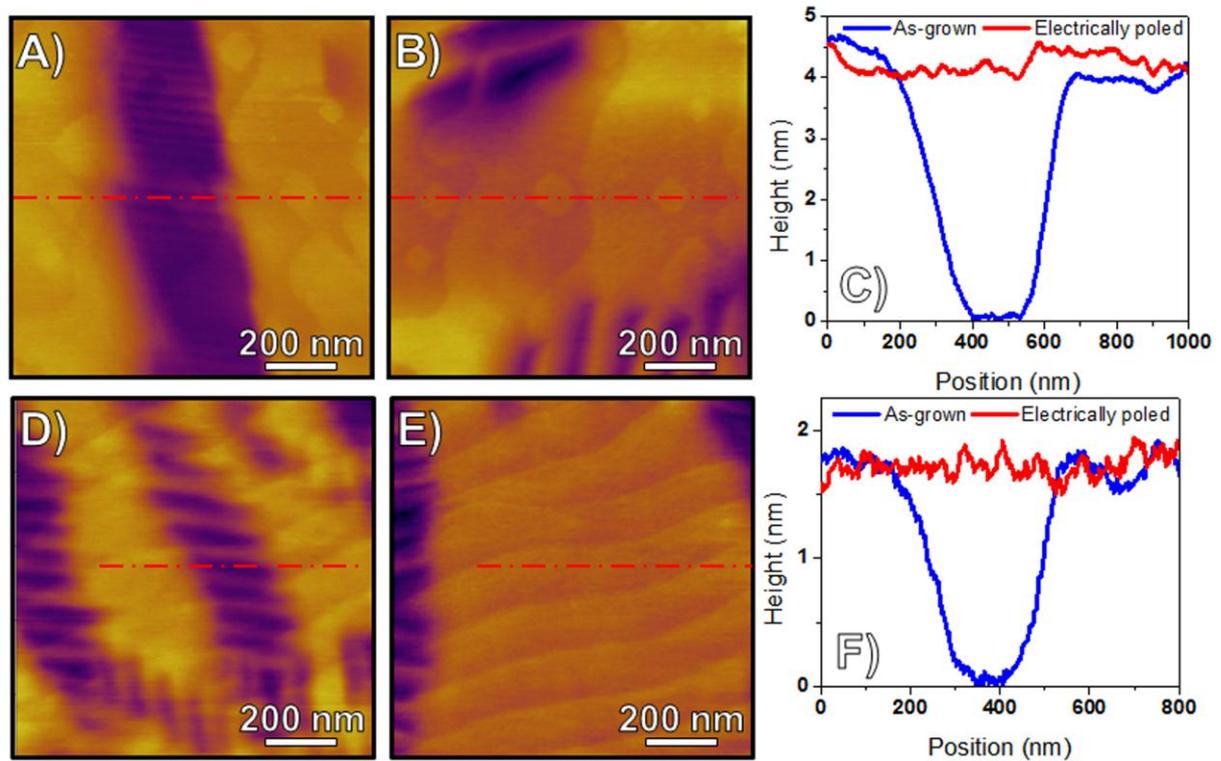

**Figure 4** AFM images of 100 nm and 50 nm BiFeO$_3$/La$_{0.5}$Sr$_{0.5}$CoO$_3$/LaAlO$_3$ (001) heterostructures in the A) and D) as-grown state, B) and E) following electrical poling, and C) and D) corresponding line traces at the red-lines.

**Figure 5**

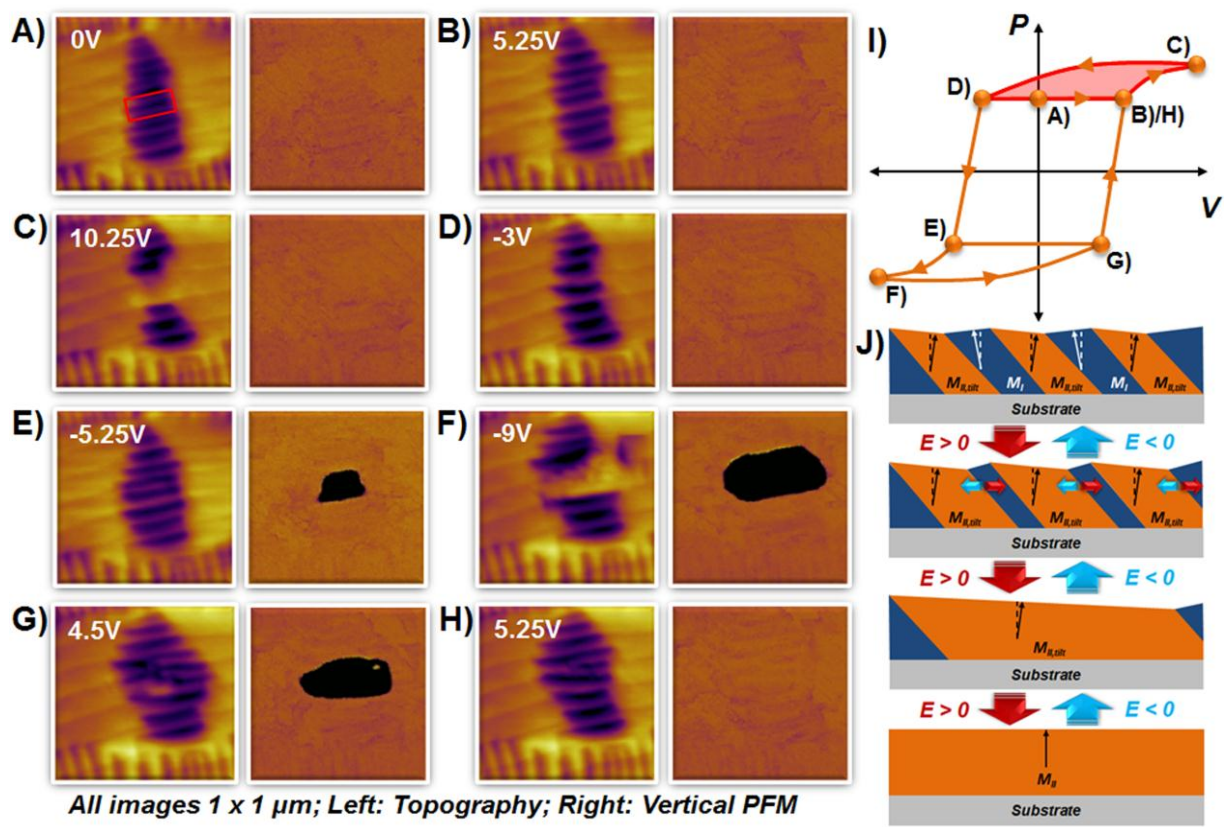

**Figure 5.** AFM image (left) and vertical PFM image (right) of 100 nm BiFeO$_3$/La$_{0.5}$Sr$_{0.5}$CoO$_3$/LaAlO$_3$ (001) in the A) as-grown state and after being poled in the red box at B) 5.25 V, C) 10.25 V, D) -3 V, E) -5.25 V, F) -9 V, G) 4.5 V, and H) 5.25 V. (All images are 1 x 1 μm). I) A schematic hysteresis loop with letters corresponding to the images in A)-H) shows the multiple pathways to enhanced electromechanical response. J) Illustration of the proposed mechanism for the large electromechanical response without the need for ferroelectric switching.